\documentclass[conf]{new-aiaa}
\usepackage{abstract}
\usepackage{simplemargins}
\usepackage{multicol}
\usepackage{datetime}
\usepackage{multirow}
\usepackage{ifthen}             
\usepackage{geometry}           
\usepackage{array}              
\usepackage{amssymb}            
\usepackage{graphicx}           
\usepackage{pstool}             
\usepackage{subfigure}          
\usepackage{fancyref}           
\usepackage{listings}           
\usepackage[noprefix]{nomencl}  
\usepackage{bm}                 
\usepackage{cancel}
\usepackage{lettrine}           
\usepackage{ifpdf}

\usepackage{lscape}
\usepackage[printonlyused]{acronym}
\usepackage{booktabs}
\graphicspath{{./figures/}}

\usepackage{bbm}
\usepackage{setspace}           

\DeclareMathOperator*{\softmax}{softmax} 

\newacro{SO}[SO]{space object}
\newacro{SSA}[SSA]{space situational awareness}
\newacro{SSN}[SSN]{Space Surveillance Network}
\newacro{SBSS}[SBSS]{Space Surveillance Network}
\newacro{UKF}[UKF]{Unscented Kalman Filter}
\newacro{EKF}[UKF]{Extended Kalman Filter}
\newacro{PF}[PF]{Particle Filter}
\newacro{GMF}[GMF]{Gaussian Mixture Filter}
\newacro{GMUKF}[GMUKF]{Gaussian Mixture Unscented Filter}
\newacro{GMEKF}[GMEKF]{Gaussian Mixture Extended Filter}
\newacro{MMAE}[MMAE]{Multiple Model Adaptive Estimation}
\newacro{SRP}[SRP]{solar radiation pressure}
\newacro{BRDF}[BRDF]{bidirectional reflectance distribution function}
\newacro{CCD}[CCD]{Charge-Coupled Device}
\newacro{LEO}[LEO]{Low Earth Orbit}
\newacro{GEO}[GEO]{Geosynchronous Orbits}
\newacro{GGQ}[GGQ]{Generalize Gaussian Quadrature}
\newacro{CGQ}[CGQ]{Classical Gaussian Quadrature}
\newacro{CGC}[CGC]{Classical Gaussian Cubature}
\newacro{GGC}[GGC]{Generalize Gaussian Cubature}
\setleftmargin{1in}
\setrightmargin{1in}
\settopmargin{1in}
\setbottommargin{1in}

\begin{document}

\title{Space Objects Maneuvering Prediction via Maximum Causal Entropy Inverse
Reinforcement Learning}
\author{Bryce Doerr\footnote{Postdoctoral Fellow,  Department of Aeronautics and Astronautics. Email: {\tt\small bdoerr@mit.edu}, AIAA Member.}}
\author{Richard Linares\footnote{Charles Stark Draper Assistant Professor, Department of Aeronautics and Astronautics. Email: {\tt\small linaresr@mit.edu}, Senior AIAA Member.}}
\affil{Massachusetts Institute of Technology, Cambridge, MA, 02139}
\author{Roberto Furfaro\footnote{Associate Professor, Department of Systems \& Industrial Engineering. Email: {\tt\small robertof@email.arizona.edu}, AIAA Member.}}
\affil{University of Arizona, Tucson, AZ, 85721}

\date{}

\maketitle{}

\begin{abstract}
Inverse Reinforcement Learning (RL) can be used to determine the behavior of Space Objects (SOs) by estimating the reward function that an SO is using for control. The approach discussed in this work can be used to analyze maneuvering of SOs from observational data. The inverse RL problem is solved using maximum causal entropy. This approach determines the optimal reward function that a SO is using while maneuvering with random disturbances by assuming that the observed trajectories are optimal with respect to the SO's own reward function. Lastly, this paper develops results for scenarios involving Low Earth Orbit (LEO) station-keeping and Geostationary Orbit (GEO) station-keeping.
\end{abstract}

\section{Introduction}
Space Situational Awareness (SSA) has many definitions depending on the goal at hand, but in general it involves collecting and maintaining knowledge of all space objects (SOs) orbiting the Earth and the space environment. This task is becoming more difficult as the number of objects currently tracked by the U.S.~increases due to breakup events and improving tracking capabilities \cite{nationalspace2010}. The Space Surveillance Network (SSN) is tasked with maintaining information on over 22,000 objects, 1,100 of which are active, with a collection of optical and radar sensors. Determining physically significant characteristics, i.e.~attributes, that go beyond simple orbital states is a key objective which is required for protecting  space capabilities and achieving SSA. For example, the SSN catalog currently includes radar cross-section and a non-conservative force parameter, analogous to a ballistic coefficient, which provides additional SO characterization information beyond position and velocity. Future SSA systems will have to be capable of building a much more detailed picture of SO attributes in order to maintain better knowledge of their characteristics, which ultimately may lead to better tracking capabilities.

This work discusses the use of inverse Reinforcement Learning (RL) to learn the behavior of Space Objects (SOs) from observed orbital motion. The behavior of SOs is estimated using inverse RL to determine the reward function that each SO is using to control. Since SOs having the capability of maneuver are controlled to achieve a particular mission-driven goal, maneuvering can be very subjective and only a data-driven learning approach can reveal the true goal. It is also important to determine what type of behavior a SO is using and if this behavior changes. Inverse RL approaches use optimal control principles to learn what reward function is being used by an agent given observations. 

The simplest inverse RL approach, discussed in Ref. \cite{abbeel2004apprenticeship}, solves for the reward function using a weighted sum of features. This is the Feature Matching Approach (FMA). The weights determined from the inverse RL calculation are the representation for the expert reward function. The estimated reward function weights can be used to determine the type of behavior mode the SO is following and to classify the model based on libraries of behavior models. These weight vectors can be added to the state of SOs as a way to represent the policy that the SO is currently following and allow for the change of this policy over time and as the behavior changes. Using the inverse RL approach, the optimal control problem is formed as a Markov Decision Process (MDP) where the reward function is not explicitly given. Rather, observations of expert demonstrations for a given task and the goal is given to estimate the reward function that the expert used to derive the demonstration trajectories. It is common to assume that the expert's actions are optimal with respect to the expert's reward function. Unfortunately, several problems exist with this early approach. First, the formulation is limited to discrete MDP system models which contain discrete state-action pairs at some time. Many problems exist with continuous state-actions (i.e. a low-thrust transfer orbit problem), thus, this formulation is limiting to just discrete MDPs. Another problem is the ambiguity between different policies and its optimality to numerous reward functions. When sub-optimal trajectories are observed, many different policies are needed to match the feature counts, and thus, many policies satisfy the feature matching. The ambiguity is not settled by this theory.   

Another earlier method to inverse RL solves for a reward function using maximum margin planning (MMP) \cite{ratliff2006maximum}. The main idea of this method is to find the state expectation that is closest to the expert demonstrations. The weights are found which form policies that have a higher expected reward than all the other policies by some margin. The demonstrated trajectories can have different feature maps, start states, and goal states which allow for more accurate reward function solutions compared to the expert. While FMA assumes that the expert is acting optimally (or close to optimally) and the reward function found by inverse RL should match the expert features closely, MMP relaxes this assumption to mimicking the expert while questioning the expert's specific reward function. Even with MMP's advantage over FMA, MMP has similar problems to FMA. Although the assumption about the expert's optimality is weakened, this assumption is still in place. Thus, for sub-optimal expert behaviors, the feature matching expectation can still be ambiguous. MMP is also formulated for discrete state and action spaces. No extension to continuous spaces has been made. 

An approach that resolves the ambiguity in the FMA and MMP assumption and has extensions to continuous state and actions is the maximum entropy formulation \cite{ziebart2008maximum}. This approach uses the principle of maximum entropy which resolves ambiguity of sub-optimal expert demonstrations to find a single stochastic policy. Specifically, maximum entropy allows for a distribution over behaviors while matching feature expectations with no duty to any specific stochastic expert trajectory. Maximum entropy has the benefit of determining solutions with a sequence of side information which are variables that are not predicted but are related to these predicted variables. If the side information is dynamic (changing through time), an extension called maximum causal entropy can be used \cite{ziebart2010modeling}. The original maximum entropy approach assumes all side information is available a priori (i.e. current known knowledge) while the side information in maximum causal entropy is revealed through time. Thus, the future side information has no causal influence on past variables. Therefore, this extension can naturally handle MDPs and state-spaces involving stochasticity and continuous state and action variables. An inverse RL framework using maximum causal entropy and a linear quadratic regulator (LQR) setting is discussed in \cite{ziebart2010modeling,ziebart2012probabilistic,monfort2015intent} and provides a solution to predicting reward functions from LQR controlled SO trajectories.

Other methods based on maximum entropy exist including nonlinear inverse RL using Gaussian processes \cite{levine2011nonlinear}, maximum entropy deep inverse RL \cite{wulfmeier2015maximum}, and generative adversarial imitation learning \cite{ho2016generative} which can solve for rewards that are complex. The main idea of inverse RL using Gaussian processes is to use a Gaussian process to express the reward function as a nonlinear function. This allows for prediction of reward functions that involve complex behaviors from sub-optimal expert trajectories. Maximum entropy deep inverse RL uses neural networks approximate complex nonlinear reward functions and sub-optimal expert trajectories. Generative adversarial imitation learning enables learning of complex expert behaviors by combining maximum causal entropy with RL techniques. By using an occupancy measure, the expert behavior is compared to the behavior learned by the RL algorithm. The expert trajectories from the SO orbit problem are formed as a quadratic reward, so these extensions are unnecessary, and a basic maximum causal entropy method is used to predict the reward function.

This work investigates the maximum causal entropy approach \cite{abbeel2004apprenticeship} for predicting the expert's reward function. Inverse RL results for Low Earth Orbit (LEO) station-keeping and Geostationary Orbit (GEO) station-keeping are presented using the maximum causal entropy. Specifically, LEO and GEO expert trajectories are formed with a LQR quadratic reward, and inverse RL using maximum causal entropy is used to predict the reward function.

\section{Reinforcement Learning}
This section provides a brief introduction to reinforcement learning. Given a discrete-time system model, we can denote the state of the system at time step $k$ by ${\bf x}_k$. The system dynamics provide the transition from ${\bf x}_k$ to ${\bf x}_{k+1}$ given ${\bf u}_{k}$, where ${\bf u}_k\in \mathcal{R}^\ell$ denotes the current control action, and this transition may be stochastic. Therefore, it is meaningful to represent this transition with a probability distribution ${\bf x}_{k+1}\sim p({\bf x}_{k+1}|{\bf x}_{k},{\bf u}_{k})$ and ${\bf x}_{k}$, ${\bf x}_{k+1}\in \mathcal{R}^n$  denotes the current and next state, respectively. The actions are modeled probabilistically and are generated by a policy ${\bf u}_k\sim \pi({\bf u}_k|{\bf x}_k)$ where the randomness in the policy can enable exploration of the policy space while also providing optimality for certain classes of control problems. 

An agent (the maneuvering SO) has a current state ${\bf x}_k\in S$ (orbital elements) at each discrete time step $k$ and chooses an action ${\bf u}_k$ according to a policy $\pi$. For the policy $\pi$, a reward signal $r_k$ is given for a transition to a new state ${\bf x}_{k+1}$. The general objective of RL is to maximize an expectation over the discounted return, $J({\bm \theta})$, given as
\begin{equation}\label{mdpreward}
J(\bm \theta)=\mathbb{E}_{\pi_{\bm \theta}}\left\{r_k+\gamma r_{k+1}+\gamma^2 r_{k+2}+\cdots\right\},
\end{equation}
where $\gamma \in [0, 1)$ is a discount factor and ${\bm \theta}$ are policy parameters. Q-learning is a popular RL method which defines a Q-function that represents the total reward or the total ``cost-to-go" for a policy $\pi$ \cite{sutton1998reinforcement}. Once the Q-function is determined, the action with the highest value or estimated total reward is taken at each time step. Therefore, the optimal policy can be determined using the optimal Q-function. The Q-function of a policy $\pi$ is
\begin{equation}
Q^{\pi}({\bf x}_{k},{\bf u}_{k})=\mathbb{E}_{\pi_{\bm \theta}}\left\{ \sum_{i=k}^{\infty} \gamma^{i-k}r_i\right\},
\end{equation}
where the function estimates the total discounted reward for policy $\pi$ from state ${\bf x}_k$ assuming that action ${\bf u}_k$ is taken and then all following actions are sampled from policy $\pi$. Q-network based methods use neural networks parameterized by $\bm \theta$ to represent $Q^{\pi}({\bf x}_{k},{\bf u}_{k}; {\bm \theta})$, but the dependency notation is dropped for simplicity \cite{sutton1998reinforcement}. Q-networks are optimized by minimizing the following loss function,
\begin{equation}
\mathcal{L}(\bm \theta)=\left(r_k + \gamma \max_{{\bf u}_{k+1}}Q^{\pi}({\bf x}_{k+1},{\bf u}_{k+1})-Q^{\pi}({\bf x}_k,{\bf u}_k)\right)^2.
\end{equation}
This equation uses the Bellman optimality condition \cite{sutton1998reinforcement} to relate $Q^{\pi}({\bf x}_{k},{\bf u}_{k})$ to $Q^{\pi}({\bf x}_{k+1},{\bf u}_{k+1})$, and this equation can be optimized using stochastic gradient descent. Given a reward function, RL can be used to determine a Q-function which is optimal with respect to this reward function. 

\section{Inverse Reinforcement Learning }
The basic principle behind inverse RL is to find the expert's reward function, $r_k({\bf x}_k,{\bf u}_k)$, that explains the expert's behavior given the observations. Reference \cite{ziebart2010modeling} introduces a maximum causal entropy method to predicting a reward function using the principle of maximum entropy which will be discussed in the following section. 
\subsection{Causal Entropy}
As discussed in the introduction, FMA and MMP are formulated by matching feature counts. This is fundamentally ill-posed because no true optimal policy will match the feature counts of a stochastic problem. Many different policies will satisfy the constraint. The principle of maximum entropy determines the least committed probability distribution to stay within the stochastic problem constraints \cite{jaynes1957information}. The causally conditioned probability is an extension that is used when side (state) information, $\mathbf{S}$, are determined after each time-step \cite{kramer1998directed}. The causally conditioned probability of an action, $\mathbf{A}$, on $\mathbf{S}$ is
\begin{equation}\label{probaovers}
    P(\mathbf{A}^N||\mathbf{S}^N)=\displaystyle\prod_{k=1}^{N}P(A_k|\mathbf{S}_{1:k},\mathbf{A}_{1:k-1}),
\end{equation}
where $N$ is the length of time, and the probability of $A_k$ is only conditioned on $\mathbf{S}_{1:k}$ and $\mathbf{A}_{1:k-1}$. The causal entropy is determined from the causally conditioned probability given by
\begin{equation}\label{causalentropy}
\begin{aligned}
    H(\mathbf{A}^N||\mathbf{S}^N)=\mathbb{E}_{P(\mathbf{A},\mathbf{S})}[-\log P(\mathbf{A}^N||\mathbf{S}^N)]\\
    =\sum_{k=1}^{N}H(A_k|\mathbf{S}_{1:k},\mathbf{A}_{1:k-1}),
\end{aligned}
\end{equation}
where $\mathbb{E}_{P(\mathbf{A},\mathbf{S})}$ is the expectation of the joint probability $P(\mathbf{A},\mathbf{S})$ and the causal entropy, $ H(\mathbf{A}^N||\mathbf{S}^N)$, is the sum of all the entropy for the causally conditioned distribution. Thus, the overall causal entropy measures the uncertainty in the causal conditioned probability distribution. If future information of the distribution is used (e.g. $\mathbf{S}_{1:N}$), the additional information decreases the causal entropy because there is less uncertainty present. The maximum causal entropy approach predicts a policy (actions) from $P(A_k|\mathbf{S}_{1:k},\mathbf{A}_{1:k-1})$ based on the state information provided, thus yielding an option to predict a reward function. Note that the joint distribution, $P(\mathbf{A},\mathbf{S})$, can be decomposed to $P(\mathbf{A},\mathbf{S})=P(\mathbf{A}^N||\mathbf{S}^N)P(\mathbf{S}^N||\mathbf{A}^{N-1})$.
\subsection{Optimization for Maximum Causal Entropy}
In order to predict an expected reward function from expert trajectories, an optimization problem is defined by maximizing the causal entropy in Eq. \eqref{causalentropy}. The causal distribution is constrained to match the expert's feature function defined by
\begin{equation}
\mathcal{F}(\mathbf{S},\mathbf{A})=\sum_k F(S_k,A_k),    
\end{equation}
which is just the sum of features through each time-step $k$. The empirical expected feature function is defined as 
\begin{equation}\label{featureexp}
\tilde{\mathbb{E}}_{\mathbf{S},\mathbf{A}}[\mathcal{F}(\mathbf{S},\mathbf{A})]=\tilde{\mathbb{E}}_{\mathbf{S},\mathbf{A}}\left[\sum_k F(S_k,A_k)\right].    
\end{equation}
The $\tilde{\mathbb{E}}_{\mathbf{S},\mathbf{A}}[\mathcal{F}(\mathbf{S},\mathbf{A})]$ statistics are usually due to limited sampling of the expert.

Then, an optimization problem can be formed as
 \begin{equation}\label{optprob}
     \begin{gathered}
     \max_{P(A_k|\mathbf{S}_{1:k},\mathbf{A}_{1:k-1})}H(\mathbf{A}^N,\mathbf{S}^N)\\
     s. t.: \mathbb{E}_{\mathbf{S},\mathbf{A}}[\mathcal{F}(\mathbf{S},\mathbf{A})]=\tilde{\mathbb{E}}_{\mathbf{S},\mathbf{A}}[\mathcal{F}(\mathbf{S},\mathbf{A})],\\
     \forall_{\mathbf{S}_{1:k},\mathbf{A}_{1:k-1}}:\hspace{12pt} \sum_{A_k}P(A_k|\mathbf{S}_{1:k},\mathbf{A}_{1:k-1})=1,\\
     \text{given:}\hspace{12pt}P(\mathbf{S}^k||\mathbf{A}^{k-1}).
     \end{gathered}
 \end{equation}
From the constraints in Eq. \eqref{optprob}, it is assumed that the state dynamics are provided explicitly with a form
\begin{equation}\label{dynamics}
    P(\mathbf{S}^k||\mathbf{A}^{k-1})=\prod_k P(S_k|A_{k-1},S_{k-1}).
\end{equation}
Therefore, a prediction on the reward function can be found through optimization of Eq. \eqref{optprob}.
\subsection{Application to MDPs and LQR}
Both MDPs and LQRs provide methods to represent stochastic systems within an optimal control structure. In inverse RL (or inverse optimal control), the goal is to predict an unknown reward function that a policy or controller is optimized about. Predicting this unknown reward function may give control policies that are close to optimal \cite{kalman1964linear}. As discussed previously, the inverse RL problem contains side information and decisions which are the states and actions respectively. This information is both dependent and stochastic. The relation is provided by the dynamics in Eq. \eqref{dynamics}.

\subsubsection{MDPs}
For MDPs, a general reward (or objective) function is given by Eq. \eqref{mdpreward} which gives a discounted reward for future steps in time. Each reward, $r_k({\bf x}_k,{\bf u}_k)$, is
\begin{equation}
    r_k({\bf x}_k,{\bf u}_k)=\theta^Tf_{{\bf x}_k,{\bf u}_k},
\end{equation}
which is a linear function between the weights, $\theta$, and the feature vector (or counts) $f_{{\bf x}_k,{\bf u}_k}$. For an MDP, the optimal policy is determined by finding an action for each state with the highest expected cumulative reward. This can occur with either finite or infinite time horizons \cite{bellman1957markovian}. The optimal policy is found by solving the the Bellman equations given by
\begin{equation}
    \begin{gathered}
    \pi(\mathbf{S}_k)=\max_{\mathbf{A}_k}\left[r_k(\mathbf{S}_k,\mathbf{A}_k)+\gamma \sum_{\mathbf{S}_{k+1}}P(\mathbf{S}_{k+1}|\mathbf{S}_k,\mathbf{A}_k)V(\mathbf{S}_{k+1})\right]\\
    V^*(\mathbf{S}_k)=\max_{\mathbf{A}_k}\left[r_k(\mathbf{S}_k,\mathbf{A}_k)+\gamma \sum_{\mathbf{S}_{k+1}}P\left(\mathbf{S}_{k+1}|\mathbf{S}_k,\mathbf{\pi}_k\left(\mathbf{S}_k\right)\right)V^*(\mathbf{S}_{k+1})\right],
    \end{gathered}
\end{equation}
which can be alternatively formulated with a value function, $V^*(\mathbf{S}_k)$, and a Q-function $Q^*(\mathbf{S}_k,\mathbf{A}_k)$ given by
\begin{equation}\label{recursive}
    \begin{gathered}
    Q^*(\mathbf{S}_k,\mathbf{A}_k)=\gamma \sum_{\mathbf{S}_{k+1}}P(\mathbf{S}_{k+1}|\mathbf{S}_k,\mathbf{A}_k)V^*(\mathbf{S}_{k+1})\\
    V^*(\mathbf{S}_k)=\max_{\mathbf{A}_k}\left[r_k(\mathbf{S}_k,\mathbf{A}_k)+Q^*(\mathbf{S}_k,\mathbf{A}_k)\right].
    \end{gathered}
\end{equation}
Equation \eqref{recursive} is recursively solved using dynamic programming and value iteration to obtain the policy $\pi(\mathbf{S}_k)$ which is similar to the previous Q-learning discussion \cite{bellman1957markovian}. For the inverse RL problem, the Q-function and the value function use a softmax function rather than a maximum function as given by
\begin{equation}
     \begin{gathered}
     Q^{\text{soft}}(\mathbf{S}_k,\mathbf{A}_k)=\gamma \sum_{\mathbf{S}_{k+1}}P(\mathbf{S}_{k+1}|\mathbf{S}_k,\mathbf{A}_k)V^*(\mathbf{S}_{k+1})\\
    V^{\text{soft}}(\mathbf{S}_k)=\softmax_{\mathbf{A}_k}\left[r_k(\mathbf{S}_k,\mathbf{A}_k)+Q^*(\mathbf{S}_k,\mathbf{A}_k)\right].
      \end{gathered}
\end{equation}
Otherwise both equations are equivalent and can be solved similarily. The softmax function allows a smooth differentiable interpolation of the maximum of functions.
\subsubsection{LQR}
A special case exists using LQR by assuming the dynamics are linear and the stochasticity is due to Gaussian noise. Thus, ${\bf u}_k\sim \pi({\bf u}_k|{\bf x}_k,\Sigma_{{\bf u}_k})$ and ${\bf x}_{k+1}\sim p(A_k{\bf x}_{k}+B_k{\bf u}_{k},\Sigma_{{\bf x}_{k+1}})$ are Gaussian.
For the classic LQR problem, the reward function, instead, is defined as a cost function, therefore, they are negatives of each other. For the finite horizon $N$, the total cost is calculated from initial state ${\bf x}_0$ and the control sequence $U=[{\bf u}_{k},{\bf u}_{k+1},\cdots,{\bf u}_{N-1}]$ applied to the dynamics given by
\begin{equation}\label{costlqr}
J({\bf x}_0,U)=\sum_{k=0}^{N-1}l({\bf x}_k,{\bf u}_k)+l_f({\bf x}_N),
\end{equation}
where $l({\bf x}_k,{\bf u}_k)$ is the running cost and $l_f({\bf x}_N)$ is the terminal cost. The LQR costs are quadratic given by
\begin{equation}\label{costatk}
l({\bf x}_k,{\bf u}_k)=\frac{1}{2}\left[ \begin{array}{c} 1 \\ { \bf x}_k \\ \mathbf{u}_k \end{array} \right]^{T} \begin{bmatrix} 0& \mathbf{q}_{k}^{T} & \mathbf{r}_{k}^{T} \\ \mathbf{q}_{k}&Q_{k} & P_{k} \\\mathbf{r}_{k}&P_{k} & R_{k}  \end{bmatrix}\left[ \begin{array}{c} 1\\ { \bf x}_k \\ \mathbf{u}_k \end{array} \right],\hspace{12pt}l_f({\bf x}_N)=\frac{1}{2}{ \bf x}_N ^{T}Q_{N}{ \bf x}_N+{ \bf x}_N ^{T}\mathbf{q}_{N},
\end{equation}
where $\mathbf{q}_{k}$, $\mathbf{r}_{k}$, $Q_{k}$, $R_{k}$, and $P_{k}$ are the running weights (coefficients) and $Q_{N}$ and $\mathbf{q}_{N}$ are the terminal weights. The weight matrices, $Q_{k}$ and $R_{k}$, are positive definite and the block matrix $\begin{bmatrix} Q_{k} & P_{k} \\P_{k} & R_{k}  \end{bmatrix}$ is positive-semidefinite \cite{inaba2016robotics}. The running and terminal costs are substituted into Eq. \eqref{costlqr}, and due to the symmetry in the weight matrices, the total cost is simplified to
\begin{equation}
J({\bf x}_0,U)=\sum_{k=0}^{N-1}{ \bf x}_k ^{T}\mathbf{q}_{k}+{ \bf u}_k^{T}\mathbf{r}_{k}+\frac{1}{2}{ \bf x}_k ^{T}Q_{k}{ \bf x}_k+\frac{1}{2}{ \bf u}_k ^{T}R_{k}{ \bf u}_k +{ \bf u}_k ^{T}P_{k}{ \bf x}_k+\frac{1}{2}{ \bf x}_N ^{T}Q_{N}{ \bf x}_N+{ \bf x}_N ^{T}\mathbf{q}_{N}.
\end{equation}
The optimal control solution is based on minimizing the cost function in terms of the control sequence which is given by
\begin{equation}\label{minJ}
U^*({\bf x}_0)=\min_U J({\bf x}_0,U).
\end{equation}
To solve for the optimal control solution given by Eq. \eqref{minJ}, a value iteration method is used. Value iteration is a method that determines the optimal cost-to-go (value) starting at the final time-step and moving backwards in time minimizing the control sequence. The cost-to-go is defined as
\begin{equation}\label{costgo}
J({\bf x}_{k},U_{k})=\sum_{k}^{N-1}l({\bf x}_k,{\bf u}_k)+l_f({\bf x}_N),
\end{equation}
where $U_{k}=[{\bf u}_{k},{\bf u}_{k+1},\cdots,{\bf u}_{N-1}]$. This is very similar to Eq. \eqref{costlqr}, but the only difference is that the cost starts from time-step $k$ instead of $k=0$. The optimal cost-to-go is calculated similar to Eq. \eqref{minJ} which is
\begin{equation}\label{valuefunction}
V({\bf x}_{k})=\min_{U_{k}} J({\bf x}_{k},U_{k}).
\end{equation}
At a time-step $k$, the optimal cost-to-go function is a quadratic function given by
\begin{equation}\label{quadfun}
V({\bf x}_{k})=\frac{1}{2}{\bf x}_k^{T}S_k{\bf x}_k+{\bf x}_k^{T}\mathbf{s}_k+ c_k,
\end{equation}
where $S_k$, $\mathbf{s}_k$, and $c_k$ are computed backwards in time using the value iteration method. First, the final conditions $S_N=Q_N$, $\mathbf{s}_N=\mathbf{q}_N$, and $c_N=c$ are set. This reduces the minimization of the entire control sequence to just a minimization over a control input at a time-step which is the principle of optimality \cite{bellman1954theory}.  To find the optimal cost-to-go, the Riccati equations are used to propagate the final conditions backwards in time given by
\begin{subequations}\label{rit}
\begin{equation}\label{rit1}
S_k=A_k^{T}S_{k+1}A_k+Q_k-\left(B_k^{T}S_{k+1}A_k+P_k^{T} \right)^{T}\left(B_k^{T}S_{k+1}B_k+R_k \right)^{-1}\left(B_k^{T}S_{k+1}A_k+P_k^{T}\right),
\end{equation}
\begin{equation}\label{rit2}
\begin{split}
\mathbf{s}_k&=\mathbf{q}_k+A_k^{T}\mathbf{s}_{k+1}+A_k^{T}S_{k+1}\mathbf{g}_k\\
&-\left(B_k^{T}S_{k+1}A_k+P_k^{T}\right)^{T}\left(B_k^{\mathsf{T}}S_{k+1}B_k+R_k\right)^{-1}\left(B_k^{T}S_{k+1}\mathbf{g}_k+B_k^{T}\mathbf{s}_{k+1}+\mathbf{r}_k\right),
\end{split}
\end{equation}
\begin{equation}\label{rit3}
\begin{split}
c_k&=\mathbf{g}_k^{T}S_{k+1}\mathbf{g}_k+2\mathbf{s}_{k+1}^{T}\mathbf{g}_k+c_{k+1}\\
&-\left(B_k^{T}S_{k+1}\mathbf{g}_k+B_k^{T}\mathbf{s}_{k+1}+\mathbf{r}_k\right)^{T}\left(B_k^{\mathsf{T}}S_{k+1}B_k+R_k\right)^{-1}\left(B_k^{T}S_{k+1}\mathbf{g}_k+B_k^{T}\mathbf{s}_{k+1}+\mathbf{r}_k\right).
\end{split}
\end{equation}
\end{subequations}

The Q-function is defined as
\begin{equation}\label{qfcnlqr}
Q({\bf x}_{k},{\bf u}_{k})=   l({\bf x}_k,{\bf u}_k)+V({\bf x}_{k+1}),
\end{equation}
for LQR \cite{rizvi2018output,sutton1998reinforcement}. By substituting Eq.\eqref{costatk} and \eqref{quadfun} into Eq. \eqref{qfcnlqr}, the Q-function has the form
\begin{equation}
    Q({\bf x}_{k},{\bf u}_{k})=\frac{1}{2}\left[ \begin{array}{c} 1 \\ { \bf x}_k \\ \mathbf{u}_k \end{array} \right]^{T} \begin{bmatrix} 0& \mathbf{q}_{k}^{T} & \mathbf{r}_{k}^{T} \\ \mathbf{q}_{k}&Q_{k} & P_{k} \\\mathbf{r}_{k}&P_{k} & R_{k}  \end{bmatrix}\left[ \begin{array}{c} 1\\ { \bf x}_k \\ \mathbf{u}_k \end{array} \right]+\frac{1}{2}{\bf x}_{k+1}^{T}S_{k+1}{\bf x}_{k+1}+{\bf x}_{k+1}^{T}\mathbf{s}_{k+1}+ c_{k+1}.
\end{equation}
From the state update equation, ${\bf x}_{k+1}=A_k{\bf x}_{k}+B_k{\bf u}_{k}$, the Q-function is reduced to its simplest form
\begin{equation}\label{qfcnlqrsim}
    \begin{split}
    Q({\bf x}_{k},{\bf u}_{k})&=\frac{1}{2}\left[ \begin{array}{c} 1 \\ { \bf x}_k \\ \mathbf{u}_k \end{array} \right]^{T} \begin{bmatrix} 0& \mathbf{q}_{k}^{T} & \mathbf{r}_{k}^{T} \\ \mathbf{q}_{k}&Q_{k} & P_{k} \\\mathbf{r}_{k}&P_{k} & R_{k}  \end{bmatrix}\left[ \begin{array}{c} 1\\ { \bf x}_k \\ \mathbf{u}_k \end{array} \right]\\
    &+\frac{1}{2}(A_k{\bf x}_{k}+B_k{\bf u}_{k})^{T}S_{k+1}(A_k{\bf x}_{k}+B_k{\bf u}_{k})\\
    &+(A_k{\bf x}_{k}+B_k{\bf u}_{k})^{T}\mathbf{s}_{k+1}+ c_{k+1}\\
    &=\frac{1}{2}\left[ \begin{array}{c} 1 \\ { \bf x}_k \\ \mathbf{u}_k \end{array} \right]^{T} \begin{bmatrix}  2c_{k+1}& \mathbf{q}_{k}^{T}+\mathbf{s}_{k+1}^TA_k & \mathbf{r}_{k}^{T}+\mathbf{s}_{k+1}^TB_k \\ \mathbf{q}_{k}+A_k^T\mathbf{s}_{k+1}&Q_{k}+A_k^TS_{k+1}A_k & P_{k}+A_k^TS_{k+1}B_k \\\mathbf{r}_{k}+B_k^T\mathbf{s}_{k+1}&P_{k}+B_k^TS_{k+1}A_k & R_{k}+B_k^TS_{k+1}B_k  \end{bmatrix}\left[ \begin{array}{c} 1\\ { \bf x}_k \\ \mathbf{u}_k \end{array} \right].
    \end{split}
\end{equation}
Using the Ricatti solution from Eq. \eqref{rit}, the optimal control policy is in the affine form
\begin{equation}
\mathbf{u}_k({\bf x}_k)=K_k{\bf x}_k+\mathbf{l}_k,
\end{equation}
where $K_k$ is the controller given by
\begin{equation}\label{controlk}
K_k=-(R_k+B_k^{T}S_{k+1}B_k)^{-1}(B_k^{T}S_{k+1}A_k+P_k^{T}),
\end{equation}
and $l_k$ is the controller offset given by
\begin{equation}\label{controloffset}
\mathbf{l}_k=-(R_k+B_k^{T}S_{k+1}B_k)^{-1}(B_k^{T}S_{k+1}\mathbf{g}_k+B_k^{T}\mathbf{s}_{k+1}+\mathbf{r}_k).
\end{equation}
\subsection{Maximum Causal Entropy with LQR}
The causal conditioned probability (or policy) in Eq. \eqref{probaovers} is necessary to determine the causal entropy in Eq. \eqref{causalentropy} which is to be maximized. For the quadratic cost function weights given in Eq. \eqref{costatk}, the value function (Eq. \eqref{quadfun}) and Q-function (Eq. \eqref{qfcnlqrsim}) can be found recursively through a value iteration method to determine the causal conditioned probability and causal entropy using 
\begin{equation}
    P(\mathbf{u}_k|\mathbf{x}_{1:k},\mathbf{u}_{1:k-1})=\exp(Q({\bf x}_{k},{\bf u}_{k})-V({\bf x}_{k})),
\end{equation}
which is Gibbs distribution \cite{monfort2016methods}. To predict the LQR weights in Eq. \eqref{costatk} from demonstrated behavior, the optimization problem in Eq. \eqref{optprob} must be solved, but the weights can be predicted through gradient-based techniques using the feature expectation in Eq. \eqref{featureexp}. The feature expectation gives the likelihood of trajectories using some action distribution. By optimizing the likelihood of predicted trajectories with the demonstrated trajectories under an action distribution, weights will be found that converges to the behavior of the demonstrated trajectories. The gradient is the difference between the expected empirical feature function and the expected predicted feature function defined by
\begin{equation}\label{gradfeature}
\begin{gathered}
    \nabla_M L=\tilde{\mathbb{E}}_{\mathbf{S},\mathbf{A}}\left[\sum_k F(S_k,A_k)\right]-\mathbb{E}_{\mathbf{S},\mathbf{A}}\left[\sum_k F(S_k,A_k)\right]\\
        =\tilde{\mathbb{E}}_{\mathbf{S},\mathbf{A}}\left[\sum_k \left[ \begin{array}{c} 1 \\ { \bf x}_k \\ \mathbf{u}_k \end{array} \right]\left[ \begin{array}{c} 1 \\ { \bf x}_k \\ \mathbf{u}_k \end{array} \right]^T\right]-\mathbb{E}_{\mathbf{S},\mathbf{A}}\left[\sum_k \left[ \begin{array}{c} 1 \\ { \bf x}_k \\ \mathbf{u}_k \end{array} \right]\left[ \begin{array}{c} 1 \\ { \bf x}_k \\ \mathbf{u}_k \end{array} \right]^T\right]
        \end{gathered}
\end{equation}
which is just the difference between the demonstrated and predicted feature vector for the LQR quadratic cost function. $M$ is the block weight matrix in Eq. \eqref{costatk}. The expectation is easily available from the feature vectors since the stochasticity in the system is Gaussian noise. A general expectation for the quadratic feature vectors given in Eq. \eqref{gradfeature} is
\begin{equation}
    E\left[\mathbf{m}\mathbf{m}^T\right]=\mu_\mathbf{m}\mu_\mathbf{m}^T+\Sigma_\mathbf{m},
\end{equation}
where $\mu_\mathbf{m}$ and $\Sigma_\mathbf{m}$ are the mean and covariance of $\mathbf{m}$ respectively. Therefore, the block weight matrix, $M$, can be iteratively optimized by a simple gradient update given by
\begin{equation}
    M=M+\eta\nabla_M,
\end{equation}
but more effective gradient methods can be used. By recursively updating the Q-function and value function for the estimated trajectories and updating the block weight matrix, $M$, iteratively through gradient decsent, a cost function that represents the expert's demonstrated trajectory can be predicted.

\section{Dynamical Models}
To show the viability of inverse RL for SOs, a orbital model is used to describe dynamics. Perturbations due to aerodynamic drag, solar radiation pressure, and $J_2$ gravitational disturbances are included in the orbtial model formulation. The nonlinear dynamic equations of motion are linearized about a reference orbit operating point for station-keeping control and are used for the cost function prediction via maximum causal entropy.
\subsection{Orbital Dynamics Model}
The dynamics for a SO orbiting Earth is described by the two-body problem is given by
\begin{equation}\label{eom}
    \ddot{\mathbf{r}}(t)=-\mu_E\frac{\mathbf{r}(t)}{r(t)^3}+\mathbf{a}_d(t),
\end{equation}
where $\mathbf{r}(t)=[x(t),\hspace{12pt}y(t),\hspace{12pt}z(t)]^T$ is the Cartesian coordinate position vector with respect to the geocentric equatorial frame, $r(t)$ is the position magnitude, $\mu_E$ is the geocentric gravitational constant, and $\mathbf{a}_d(t)$ are perturbing accelerations. The perturbing accelerations, $\mathbf{a}_d(t)$, can be expanded to contain external disturbances given by
\begin{equation}
    \mathbf{a}_d(t)=\mathbf{a}_c(t)+\mathbf{a}_{ad(t)}+\mathbf{a}_{J_2}(t)+\mathbf{a}_{srp}(t),
\end{equation}
where $\mathbf{a}_c(t)$ is control by thrusters, $\mathbf{a}_{ad}(t)$ is aerodynamic drag, $\mathbf{a}_{J_2}(t)$ is gravitational perturbations, and $\mathbf{a}_{srp}(t)$ is solar radiation pressure. 

\subsubsection{Control by Thrust}
To control a SO, delta-v maneuvers take place over time. An external force from the thrusters is produced which allows the SO to actively control against any perturbations for station-keeping. The acceleration due to a thrust, $T$, is 
\begin{equation}
    \mathbf{a}_c(t)=\frac{T(t)\dot{\mathbf{r}}(t)}{m_{SO}(t)||\dot{\mathbf{r}}(t)||}
\end{equation}
where $m_{SO}(t)$ is the mass of the SO, $\dot{\mathbf{r}}(t)$ is the velocity vector, and $||\cdot||$ is the $L_2$ norm. This shows that the accelerations obtained from a thrust, $T(t)$, must coincide with the direction of the velocity vector. For chemical thrusters, the SO mass decreases due to expenditure of the propellant in the rocket. Therefore, the SO mass decreases in time and can be described by a first order ordinary differential equation,
\begin{equation}
    \dot{m_f}(t)=\frac{T(t)}{I_{sp}g_0},
\end{equation}
where $I_{sp}$ is the propulsion system's specific impulse, $g_0$ is the standard gravity at sea-level, and $m_f(t)$ is the mass of the propellant. Thus, the total SO mass is $m_{SO}(t)=m_f(t)+m_p$, where $m_p$ is the payload mass of the SO. 
\subsubsection{General Dynamical System}
The two-body problem and the propellant mass loss rate are dependent, so to propagate the orbital dynamics and mass through time, the state vector and the nonlinear dynamical system are defined as
\begin{equation}\label{statevector}
 \mathbf{x}(t)= \begin{bmatrix}x(t)\\ y (t)\\ z(t) \\\dot{x}(t)\\  \dot{y}(t) \\\dot{z}(t)\\m_f(t) \end{bmatrix}   , \hspace{12pt} \dot{\mathbf{x}}=f(t,\mathbf{x}(t),\mathbf{a}_d(t))=\begin{bmatrix}\dot{x}(t)\\ \dot{y}(t) \\ \dot{z}(t)\\-\mu_E\frac{x(t)}{r(t)^3}+\mathbf{a}_{d_x}(t)\\ -\mu_E\frac{y(t)}{r(t)^3}+\mathbf{a}_{d_y}(t) \\-\mu_E\frac{z(t)}{r(t)^3}+\mathbf{a}_{d_z}(t)\\\frac{T(t)}{I_{sp}g_0} \end{bmatrix}.
\end{equation}
The time-step $t$ is continuous and both $\mathbf{x}(t)$ and $\mathbf{a}_d(t)$ vary with time. For simplicity of notation, the dependency of the time-step, $(t)$, is removed. Solutions of the state through time can be solved with numerical solvers (e.g. a Runge–Kutta 4th order method).
\subsection{Aerodynamic Drag}
Aerodynamic drag is a disturbance that affects the life-time of the spacecraft. This drag can disturb the SO from its nominal orbit and possibly deorbit the SO without any correctional maneuver. Since fuel used to correct the trajectory from the disturbance is finite, aerodynamic drag is a disturbance that limits the life-span of the spacecraft. The disturbance becomes more apparent for orbits that are smaller in altitude. Therefore, a SO that is situated in a LEO orbit has more aerodynamic drag than a SO in a GEO orbit. Many models exist that relate the atmospheric density with the altitude on Earth \cite{johnson2002reference}. For this paper, the US Standard Atmosphere 1976 (USSA76) \cite{atmosphere1976national} is used to model the aerodynamic drag that occurs on the SO while in orbit. The USSA76 is assumed to be a steady-state $1000 km$, spherically symmetric gaseous shell surrounding Earth. The variations in the density with altitude is approximated by averaging the year-round conditions at mid-latitudes over many years. Therefore, the model gives realistic densities generally that may not coincide with actual values at a specific point.

To find the disturbance, $\mathbf{a}_{ad}$, the aerodynamic drag force, $\mathbf{D}_{ad}$, acts opposite of the SO velocity relative to the atmosphere, $\hat{\dot{\mathbf{r}}}_{rel}$, given by
\begin{equation}
    \mathbf{D}_{ad}=-D\hat{\dot{\mathbf{r}}}_{rel},
\end{equation}
where $\dot{\mathbf{r}}_{rel}$ is
\begin{equation}
    \begin{split}
    \dot{\mathbf{r}}_{rel}&=\dot{\mathbf{r}}-\dot{\mathbf{r}}_{atm}\\
    &=\dot{\mathbf{r}}-\omega_E\times\mathbf{r}.
    \end{split}
\end{equation}
The rotational rate of Earth is $\omega_E$, and the direction of the relative velocity is a unit vector
\begin{equation}
    \hat{\dot{\mathbf{r}}}_{rel}=\frac{\dot{\mathbf{r}}_{rel}}{||\dot{\mathbf{r}}_{rel}||}.
\end{equation}
The drag is computed as
\begin{equation}
    D=\frac{1}{2}\rho||\dot{\mathbf{r}}_{rel}||^2C_DA,
\end{equation}
where $\rho$ is the density of the atmosphere, $C_D$ is the coefficient of drag, $A$ is the area of the spacecraft. Therefore, the acceleration due to aerodynamic drag is
\begin{equation}
    \mathbf{a}_{ad}=-\frac{1}{2}\rho||\dot{\mathbf{r}}_{rel}||^2\left( \frac{C_DA}{m_{SO}}\right)\dot{\mathbf{r}}_{rel}.
\end{equation}
\subsection{Gravitational Perturbations}
It is typically assumed that the Earth has a spherically symmetric mass distribution in which the gravitational field is spherically symmetric about the center of the sphere. In reality, the Earth and other planets are not perfect spheres, but they resemble oblate spheroids instead. Thus, the gravitational field varies with the latitude and radius about Earth \cite{vallado2007_}. Still, the gravity potential is mostly contributed by the spherically symmetric assumption given by
\begin{equation}
    V_{ss}=-\frac{\mu_E}{r},
\end{equation}
where $V_{ss}$ is the gravity potential due to a spherically symmetric mass distribution. The gravitational potential perturbations due to Earth's oblateness is 
\begin{equation}\label{pertjn}
    \Phi(r,\phi)=\frac{\mu_E}{r}\sum_{k=2}^{\infty} J_k\left(\frac{R}{r}\right)^kP_k\left(\cos\phi\right),
\end{equation}
where $R$ is the equatorial radius of Earth, $J_k$ are the Earth zonal harmonics, $\phi$ is the polar angle defined by
\begin{equation}
    \phi=\arctan\left( \frac{\sqrt{x^2+y^2}}{z}\right),
\end{equation}
and $P_k$ are the Legendre polynomials. The $J_k$ are constants that are derived due to orbital motion of SO around Earth. Note that the zonal harmonic for $J_1=0$, thus, Earth has a spherical symmetric mass distribution. The dominant gravitational perturbation from Earth's oblateness is due to the $J_2=0.00108263$ zonal harmonic. Higher order zonal harmonic perturbations can be found in Reference \cite{vallado2007_}. The gravitational potential includes a Legendre polynomial term obtained by the Rodrigues' formula about an arbitrary variable, $w$, given by
\begin{equation}\label{rod}
   P_k(w)=\frac{1}{2^kk!}\frac{d}{dw^k}(w^2-1)^k, 
\end{equation}
which can be expanded to higher order terms with additional zonal harmonics \cite{vallado2007_}. By using the most dominant perturbation term, $J_2$, only the $P_2$ term is necessary for the gravitational disturbance formulation, but additional zonal harmonic terms can be added, if necessary. The $P_2$ derived from the Rodriques' formula is
\begin{equation}\label{rod2}
    P_2(w)=\frac{1}{2}(3w^2-1).
\end{equation}
By substituting Eq. \eqref{rod2} into Eq. \eqref{pertjn} for $k=2$, the gravitation potential perturbation is
\begin{equation}
    \Phi(r,\phi)=\frac{J_2\mu_E}{2r}\left(\frac{R}{r}\right)^2\left(3\cos^2\phi-1\right).
\end{equation}
The acceleration due to Earth's oblateness is defined by
\begin{equation}
    \mathbf{a}_{J_2}=-\nabla_{\mathbf{r}}\Phi(r,\phi),
\end{equation}
which is the negative gradient of $\Phi(r,\phi)$ in terms of ${\mathbf{r}}$ in the geocentric frame. After taking the gradient, $\mathbf{a}_{J_2}$ is simplified to
\begin{equation}
    \mathbf{a}_{J_2}=\frac{3J_2\mu_ER^2}{2r^4}\left[ \frac{x}{r}\left( 5\frac{z^2}{r^2}-1\right),\hspace{12pt}\frac{y}{r}\left( 5\frac{z^2}{r^2}-1\right),\hspace{12pt}\frac{z}{r}\left( 5\frac{z^2}{r^2}-3\right)\right]^T.
\end{equation}
\subsection{Solar Radiation Pressure}
Solar radiation pressure is a electronmagnetic radiation disturbance from the Sun that affects the SO in orbit. The photosphere (visible surface) of the Sun emits radiation intensity given by 
\begin{equation}
    S=S_0\left( \frac{R_0}{R_{se}}\right)^2,
\end{equation}
where $S_0$ is the Sun power intensity, $R_0$ is the radius of the photosphere, and $R_{SE}$ is the distance between the Sun and Earth. Therefore, the solar radiation intensity at Earth is $S=1367 \text{W}/\text{m}^2$ for the Sun power intensity, $63.15\times10^6\text{W}/\text{m}^2$. The solar radiation pressure for an Earth-orbiting SO is simply
\begin{equation}
    P_{srp}=\frac{S}{c},
\end{equation}
where $c$ is the speed of light. To simplify the geometry of the SO as it is disturbed by solar radiation pressure, it is assumed that the cross-sectional geometry of the SO facing the Sun is a cannonball model (i.e. $A_{srp}=\pi R_{SO}^2$ where $R_{SO}$ is the radius of the SO). The acceleration disturbance is given by
\begin{equation}\label{srpdis}
    \mathbf{a}_{srp}=-\nu\frac{P_{srp}C_RA_{srp}}{m_{SO}}\hat{\mathbf{u}},
\end{equation}
where the unit vector $\hat{\mathbf{u}}$ points to the Sun from the SO. Therefore, $\mathbf{a}_{srp}$ is negative because the solar radiation pressure points away from the Sun. The shadow function, $\nu$, is set to 0 or 1 depending on whether the SO is in Earth's shadow or not, respectively. Reference \cite{curtis2013orbital} gives an algorithm to determine whether the SO is in Earth's shadow or not. The radiation pressure coefficient, $C_R$, depends on the color of the spacecraft and lies between 1 and 2. If the cross-sectional area is a black body, all the radiation momentum is absorbed which gives $C_R=1$. If the color of the cross-sectional area is fully reflective, $C_R=2$ because all the incoming radiation momentum is reflected which doubles the force on the SO. Equation \eqref{srpdis} specifically depends on the $A_{srp}/m_{SO}$ ratio. Solar sails that have large areas and very little mass are affected more by solar radiation pressure disturbances. To simplify the $\hat{\mathbf{u}}$ calculation, it is assumed that the unit vector points from Earth to the sun instead of the SO to the Sun. This assumption is valid because the angle between the Earth-to-Sun line and the SO-to-Sun line is less than $0.02^{\circ}$ due to the large distance the Sun is away from Earth \cite{curtis2013orbital}. The, $\hat{\mathbf{u}}$ defined as
\begin{equation}
    \hat{\mathbf{u}}=\cos\lambda\hat{\mathbf{I}}'+\sin\lambda\hat{\mathbf{J}}',
\end{equation}
where $\lambda$ is the solar ecliptic longitude, and the unit vectors, $\hat{\mathbf{I}}'$ and $\hat{\mathbf{J}}'$, are part of the geocentric ecliptic frame. A simple rotation matrix involving the angle between Earth's equatorial plan and the ecliptic plane, $\epsilon$, is
\begin{equation}
    \hat{\mathbf{u}}_{XYZ}=\mathbf{R}_1(-\epsilon)\hat{\mathbf{u}}_{X'Y'Z'}
\end{equation}
which converts the geocentric ecliptic frame to the geocentric equatorial frame. Thus, the solar radiation pressure in the geocentric equatorial frame is
\begin{equation}
   \mathbf{a}_{srp}= -\nu\frac{P_{srp}C_RA_{srp}}{m_{SO}} \begin{bmatrix}\cos\lambda\\ \cos\epsilon\sin\lambda \\\sin\epsilon\sin\lambda \end{bmatrix}.
\end{equation}
To find the solar ecliptic longitude $\lambda$, the obliquity, $\epsilon$, and the Julian day since the year 2000 are necessary \cite{nautical2008astronomical}. This equation is
\begin{equation}
    n=JD-2451545,
\end{equation}
where $JD$ is the Julian day to be simulated. The obliquity, $\epsilon$, is determined directly given by
\begin{equation}
    \epsilon=23.439^{\circ}-3.56\times10^{-7}n.
\end{equation}
To determine the solar ecliptic longitude, both the mean longitude and mean anomaly of the Sun must be found which are
\begin{equation}
    L=280.459^{\circ}+0.98564736^{\circ}n \hspace{12pt}(0^{\circ}\leq L\leq360^{\circ}),
\end{equation}
\begin{equation}
    M=357.529^{\circ}+0.98560023^{\circ}n \hspace{12pt}(0^{\circ}\leq M\leq360^{\circ}),
\end{equation}
from Reference \cite{nautical2008astronomical}. Thus, the solar ecliptic longitude is
\begin{equation}
    \lambda=L+1.915^{\circ}\sin M +0.0200^{\circ}\sin 2M \hspace{12pt}(0^{\circ}\leq \lambda\leq360^{\circ}).
\end{equation}

\subsection{Linearization and Discretization}
\subsubsection{Linearization}
The dynamics for a SO orbiting Earth is described by the nonlinear equations of motion given by Eq. \eqref{eom}. The disturbing accelerations (e.g. aerodynamic drag, gravitational perturbations, and solar radiation pressure) can be modeled to high levels of accuracy for time into the future, which allows the nonlinear dynamics to be described as a linear time-varying (LTV) system. The disturbing accelerations are driven through a time-dependent functions discussed previously. By taking a first-order Taylor series expansion of Eq. \eqref{eom} around an operating point for linearization, $\bar{\mathbf{x}}$ and $\bar{\mathbf{a}}_c$, where $\bar{\mathbf{x}}$ is the predefined nominal geostationary slot for the SO (i.e. $\bar{\mathbf{x}}=\mathbf{x}_{geo}$), a linear time-varying system is found given by
\begin{equation}
    \dot{\mathbf{x}}\approx f(\mathbf{x},\mathbf{a}_c)\vert_{\bar{\mathbf{x}},\bar{\mathbf{a}}_c}+\left.\frac{\partial f}{\partial\mathbf{x}}\right\vert_{\bar{\mathbf{x}}}\left( \mathbf{x}-\bar{\mathbf{x}}\right)+\left.\frac{\partial f}{\partial\mathbf{a}_c}\right\vert_{\bar{\mathbf{a}}_c}\left( \mathbf{a}_c-\bar{\mathbf{a}}_c\right).
\end{equation}
This equation is simplified to a linear state-space equations using perturbations from the operating point given by
\begin{equation}\label{partiallinsys}
    \partial\dot{\mathbf{x}}=A\partial\mathbf{x}+B\partial\mathbf{a}_c,
\end{equation}
where $\partial\dot{\mathbf{x}}=\dot{\mathbf{x}}- f(t,\mathbf{x},\mathbf{a}_c)\vert_{\bar{\mathbf{x}},\bar{\mathbf{a}}_c}$, $\partial\mathbf{x}=\left( \mathbf{x}-\bar{\mathbf{x}}\right)$, and $\partial\mathbf{a}_c=\left( \mathbf{a}_c-\bar{\mathbf{a}}_c\right)$. The linearized $A$ is time-dependent while the $B$ matrix is time-independent. They are both given by
\begin{equation}\label{abcontinuous}
    A= \left.\begin{bmatrix}0&0&0&1&0&0&0\\ 0&0&0&0&1&0&0\\ 0&0&0&0&0&1&0 \\\frac{\mu(2x^2-y^2-z^2)}{(x^2+y^2+z^2)^{5/2}}&\frac{3\mu xy}{(x^2+y^2+z^2)^{5/2}}&\frac{3\mu xz}{(x^2+y^2+z^2)^{5/2}}&0&0&0&0\\  \frac{3\mu xy}{(x^2+y^2+z^2)^{5/2}}&\frac{\mu(-x^2+2y^2-z^2)}{(x^2+y^2+z^2)^{5/2}}&\frac{3\mu yz}{(x^2+y^2+z^2)^{5/2}}&0&0&0&0 \\\frac{3\mu xz}{(x^2+y^2+z^2)^{5/2}}&\frac{3\mu yz}{(x^2+y^2+z^2)^{5/2}}&\frac{\mu(-x^2-y^2+2z^2)}{(x^2+y^2+z^2)^{5/2}}&0&0&0&0\\0&0&0&0&0&0&0 \end{bmatrix}\right\vert_{\bar{\mathbf{x}}}   , \hspace{12pt} B=\begin{bmatrix}0&0&0\\ 0&0&0 \\ 0&0&0\\1&0&0\\ 0&1&0 \\0&0&1\\0&0&0 \end{bmatrix}.
\end{equation}
This provides an approximate linearization of the nonlinear dynamics about the geostationary slot operating point.
\subsubsection{Discretization}
The $A$ and $B$ matrices are discretized using a Runge-Kutta fourth-order with a zero-order hold on the control input $\mathbf{a}_c$ \cite{decarlo1989linear,van1978computing}. A general Runge-Kutta fourth-order integration is given by
\begin{equation}\label{4zoh}
    \mathbf{x}_{k+1}=\mathbf{x}_{k}+\frac{1}{6}h(k_1+k_2+k_3+k_4).
\end{equation}
Note that the variable $k$ is the discretized time-step and $h$ is the step size for the system. With the zero-order hold on a control input $\mathbf{u}$, the $k_1$, $k_2$, $k_3$, and $k_4$ are given by
\begin{equation}\label{kkkk}
    \begin{split}
        k_1&=h\left( A\mathbf{x}_k+B\mathbf{u}_k\right)\\
        k_2&=h\left( A\left(\mathbf{x}_k+k_1/2\right)+B\mathbf{u}_k\right)\\
        k_3&=h\left( A\left(\mathbf{x}_k+k_2/2\right)+B\mathbf{u}_k\right)\\
        k_4&=h\left( A\left(\mathbf{x}_k+k_3\right)+B\mathbf{u}_k\right).
    \end{split}
\end{equation}
By substituting Eq. \eqref{kkkk} into Eq. \eqref{4zoh}, the discretized system from a Runge-Kutta fourth-order with a zero-order hold on control is
\begin{equation}\label{discretization}
    \begin{gathered}
    \mathbf{x}_{k+1}=\left( I+hA+\frac{h^2}{2!}A^2+\frac{h^3}{3!}A^3+\frac{h^4}{4!}A^4\right)\mathbf{x}_{k}+\left( h+\frac{h^2}{2}A+\frac{h^3}{3!}A^2+\frac{h^4}{4!}A^3\right)B\mathbf{u}_k\\
    =A_k\mathbf{x}_{k}+B_k\mathbf{u}_k.
    \end{gathered}
\end{equation}
The $A_k$ is the discrete time-varying version of a continuous time-varying $A=A(t)$. Otherwise, for a continuous time-invariant $A(t+1)=A(t)$, the discretized version will also be time-invariant (i.e. $A_{k+1}=A_{k}$). Therefore, the linear system approximation given by Eq. \eqref{partiallinsys} can be discretized at each time-step, $k$, using Eq. \eqref{discretization}. Although the continuous time linearized system for the orbital equations of motion is time-dependent in $A$ and time-independent in $B$ in Eq. \eqref{abcontinuous}, the discrete time version of the same system contains a time-dependent $A_k$ and $B_k$. 
\subsection{Conversion of Cartesian Coordinates to Orbital Elements}
To describe a SO in orbital motion, either the state vector representation in Eq. \eqref{statevector} or an orbital element representation can be used. The state vector representation describes the position and velocity of SO in orbital motion which is equivalent to describing the shape and orientation of the SO in orbit using the classical orbital elements. The six orbital elements that describe the SO in orbital motion are $h$, $i$, $\Omega$, $e$, $\omega$, and $\theta$ which are the specific angular momentum, inclination, right ascension of the ascending node, eccentricity, argument of perigee, and true anomaly, respectively. Although $h$ and $\theta$ are used, they can be replaced by the semimajor axis, $a$, and the mean anomaly, $M$. The angular momentum, $h$, and the eccentricity, $e$, describe the shape of the orbit while the inclination, $i$, right ascension of the ascending node, $\Omega$, and argument of perigee, $\omega$, describe the orientation of the orbit. The true anomaly, $\theta$, describes the position of the SO on its orbit. More specific descriptions of orbital elements can be found in \cite{curtis2013orbital,gazzino2017dynamics}. The advantages of using orbital elements is that it shows the explicit shape and orientation of the orbit compared to knowing the position and velocity using Cartesian coordinates at a time-step. Thus, it is easier to determine whether the SO orbit is on an escape trajectory or orbiting at the equatorial plane using orbital elements.

The transformation of Cartesian coordinates in a geocentric equatorial frame to orbital elements is found in \cite{curtis2013orbital}. The angular momentum vector is given by
\begin{equation}\label{angmom}
    \mathbf{h}=\mathbf{r}\times\dot{\mathbf{r}}.
\end{equation}
The angular momentum orbital element, $h$, is just the $L_2$ norm of Eq. \eqref{angmom}. The inclination angle lies in the between $0^{\circ}$ and $180^{\circ}$ and is given by
\begin{equation}
    i=\arccos{\frac{h_Z}{h}},
\end{equation}
where $h_Z$ is the z coordinate of the angular momentum vector $\mathbf{h}$. To obtain the right ascension of the ascending node, the node line must be calculated. The node line is calculated as
\begin{equation}\label{nodeline}
    \mathbf{N}=\hat{K}\times\mathbf{h},
\end{equation}
where $\hat{K}$ is the unit vector in the z direction of a Cartesian system. The node magnitude, $N$, is simply the $L_2$ norm of Eq. \eqref{nodeline}. The right ascension of the ascending node is then
\begin{equation}
\begin{gathered}
    \Omega=\arccos{\frac{N_X}{N}}\hspace{12pt} (N_Y\geq0)\\
    \Omega=2\pi-\arccos{\frac{N_X}{N}}\hspace{12pt} (N_Y<0),
    \end{gathered}
\end{equation}
which resolves the quadrant ambiguity of the arccosine function from $0^{\circ}$ to $360^{\circ}$. Next, the eccentricity vector defined as
\begin{equation}
    \mathbf{e}=\frac{1}{\mu}\left[ \left( \dot{r}^2-\frac{\mu}{r}\right)\mathbf{r} -rv_r\dot{\mathbf{r}}\right],
\end{equation}
where $v_r$ is the radial velocity magnitude given by
\begin{equation}
    v_r=\frac{\mathbf{r}\cdot\dot{\mathbf{r}}}{r}.
\end{equation}
Again, the eccentricity magnitude $e$ is the $L_2$ norm. The argument of perigee is
\begin{equation}
\begin{gathered}
    \omega=\arccos\left\{\frac{\mathbf{N}\cdot\mathbf{e}}{Ne}\right\} \hspace{12pt}(e_Z\geq0)\\
    \omega=2\pi-\arccos\left\{\frac{\mathbf{N}\cdot\mathbf{e}}{Ne}\right\} \hspace{12pt}(e_Z<0),
    \end{gathered}
\end{equation}
considering the quadrant ambiguity of the arccosine function from $0^{\circ}$ to $360^{\circ}$. Lastly, the true anomaly can be solved using
\begin{equation}
\begin{gathered}
    \theta=\arccos\left\{\frac{\mathbf{e}\cdot\mathbf{r}}{er}\right\} \hspace{12pt}(v_r\geq0)\\
    \theta=2\pi-\arccos\left\{\frac{\mathbf{e}\cdot\mathbf{r}}{er}\right\} \hspace{12pt}(v_r<0),
    \end{gathered}
\end{equation}
taking into account the arccosine ambiguity from $0^{\circ}$ to $360^{\circ}$. Two singular cases can arise using orbital elements. When the orbit is circular, $e=0$ and $\omega$ is undefined. To alleviate this problem, $\omega$ is set to 0 when the node magnitude, $N=0$. When the orbit lies on the equatorial plane, $i=0$ and $\Omega$ is undefined. Similar to the other case, $\Omega$ is set to 0 when the node magnitude, $N=0$. Thus, the six orbital elements are able to alternatively describe the SO orbit based on the position and velocity in Cartesian coordinates based on the geocentric equatorial frame.
\section{Simulation Results}
This section discusses the initial proof-of-concept results for the proposed maximum causal entropy approach to learning the SO's behavior in orbit. Three cases are considered in this section. The first case discusses the inverse RL approach for GEO station-keeping. The next case applies this same approach to a LEO orbit. 
The inverse reinforcement learning approach is used to learn the reward function the produces the SO's maneuver. The learned reward function can then be used to estimate the behavior of SOs.   

  \begin{table}[htbp]
   \caption{SO Parameters} \label{SOparam}
   \small 
   \centering 
   \begin{tabular}{lr} 
   \toprule[\heavyrulewidth]\toprule[\heavyrulewidth]
   \textbf{Parameter} & \textbf{Value} \\ 
   \midrule
   Mass  & 100 kg \\
   Surface Area & 3.14 m$^2$\\
   Solar Reflection Coefficient & 1.0\\
   Specific Impulse & 300 s\\
   Maximum Thrust & 5 N\\
   Aerodynamic Drag Coefficient & 2.2\\
   \bottomrule[\heavyrulewidth] 
   \end{tabular}
\end{table}

 \subsection{GEO Results}
 
\begin{figure}[!htb]
\begin{centering}
    \subfigure[Trajectory using True LQR weights]{
        \includegraphics[keepaspectratio, width=0.45\textwidth]{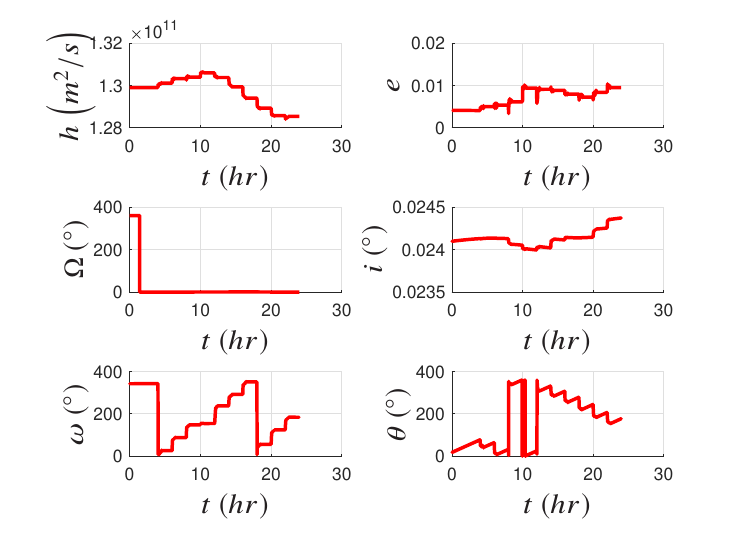}
\label{experttraj}}
   \subfigure[Trajectory using Estimated LQR weights]{
       \includegraphics[keepaspectratio, width=0.45\textwidth]{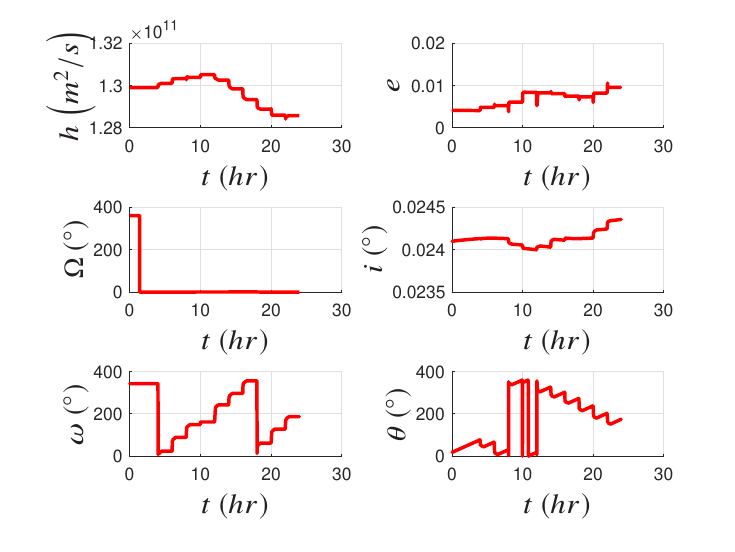}
\label{esttraj}}
    \caption{SO Trajectory using True and Estimated LQR Weights for a GEO orbit}\label{trajplotgeo}
\end{centering}
\end{figure}

\begin{figure}[!htb]
\begin{centering}
    \subfigure[True Cost Function]{
    \includegraphics[keepaspectratio, width=0.45\textwidth]{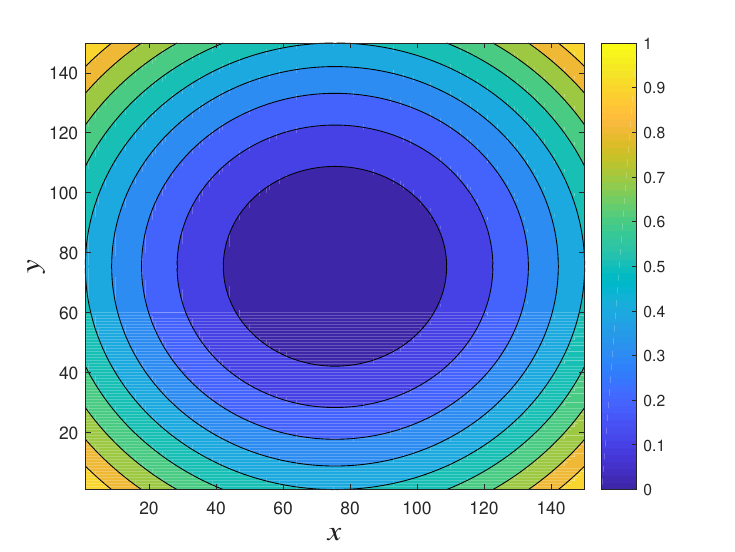}
\label{expertcon}}
\subfigure[Estimated Cost Function]{
\includegraphics[keepaspectratio, width=0.45\textwidth]{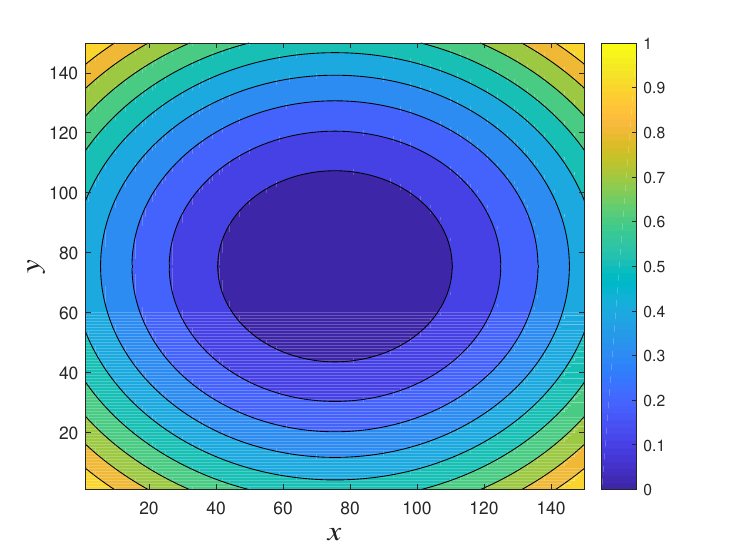}
\label{estcon}}  
\subfigure[Estimated Cost Function Error]{
\includegraphics[keepaspectratio, width=0.45\textwidth]{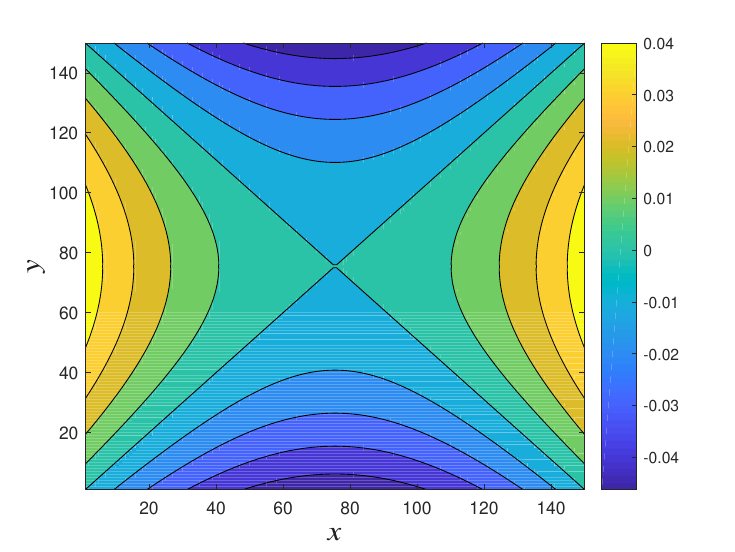}
\label{errorcon}}  
        \caption{Estimated vs. True Cost Function for GEO Maneuvering SO} \label{fig:case35des}
\end{centering}
\end{figure}

 The first case involves the simulation of a SO in a GEO orbit. In this scenario, the objective of the SO is to stay in GEO orbit under the influence of solar radiation pressure, gravitational perturbations, and aerodynamic drag. The parameters for the SO are given in Table \ref{SOparam}. The ephemeris for a SO in an initial GEO orbit is given in Table \ref{SOephGEO}. To maintain orbital station-keeping, the thrusters are fired every two hours for up to 30 minutes if the SO lies outside a 75km box of the nominal GEO orbit. This is to provide a conventional thruster burn schedule for station-keeping maneuvers \cite{de2015geostationary}. To obtain the expert trajectories, a hundred trajectories were simulated with Gaussian white noise. The cost function for these trajectories were determined using the quadratic cost function in Eq. \eqref{costatk} but with constant weight matrices through time. Figure \ref{experttraj} shows a single expert trajectory in GEO orbit with an initial perturbation from the nominal GEO orbit. From this figure, the eccentricity, inclination, and right ascension of the ascending node stay approximately zero through time due to stabilizing LQR control to the nominal GEO orbit. This figure explicitly shows the delta-v burns that occur every two hours when the SO is outside the 75km box around the nominal GEO orbit. A very small transient response in each element can be observed due to the burn itself.  This control schedule is able to mitigate the variances due to the disturbances in the system. Figure \ref{expertcon} shows the true cost that is used for every expert trajectory. This LQR cost, which is what is considered to be unknown, is estimated using maximum causal entropy inverse RL.
 
 \begin{table}[htbp]
   \caption{SO ephemeris at an initial GEO orbit} \label{SOephGEO}
   \small 
   \centering 
   \begin{tabular}{lr} 
   \toprule[\heavyrulewidth]\toprule[\heavyrulewidth]
   \textbf{Parameter} & \textbf{Value} \\ 
   \midrule
   Semimajor Axis  & 42166.7 km \\
   Eccentricity & 0\\
   Inclination & 0$^{\circ}$\\
   Right Ascension of the Ascending Node & 0$^{\circ}$\\
   Argument of Perigee & 0$^{\circ}$\\
   True Anomaly & 0$^{\circ}$\\
   Epoch & July, 1st, 2020 $00:00:00.0$ UT1\\
   \bottomrule[\heavyrulewidth] 
   \end{tabular}
\end{table}

 Figure \ref{esttraj} shows a trajectory which uses the estimated LQR weights determined by maximum causal entropy. The trajectory is initialized with the same initial conditions as Figure \ref{experttraj}. Similar to the expert trajectory, the SO is able to mitigate disturbances and maintain a GEO orbit with the same delta-v control schedule. The time-history is similar, but Figure \ref{esttraj} has a less distinct transient response from the delta-v burns. The estimated cost function is shown in Figure \ref{estcon}. The estimated cost function is able to match the behavior of the expert trajectory. Specifically, the maximum error between the true and estimated cost function is 0.04 as shown in Figure \ref{errorcon}. Even with these small differences, the estimated cost function captures the general trend of the station-keeping maneuver.

 \subsection{LEO Results}
\begin{figure}[!htb]
\begin{centering}
    \subfigure[Trajectory using True LQR weights]{
    \includegraphics[keepaspectratio, width=0.45\textwidth]{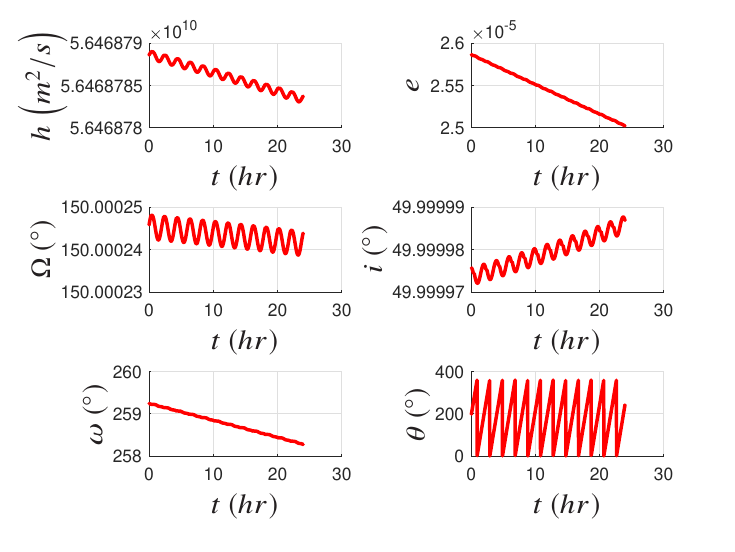}
\label{experttrajleo}}
   \subfigure[Trajectory using Estimated LQR weights]{
       \includegraphics[keepaspectratio, width=0.45\textwidth]{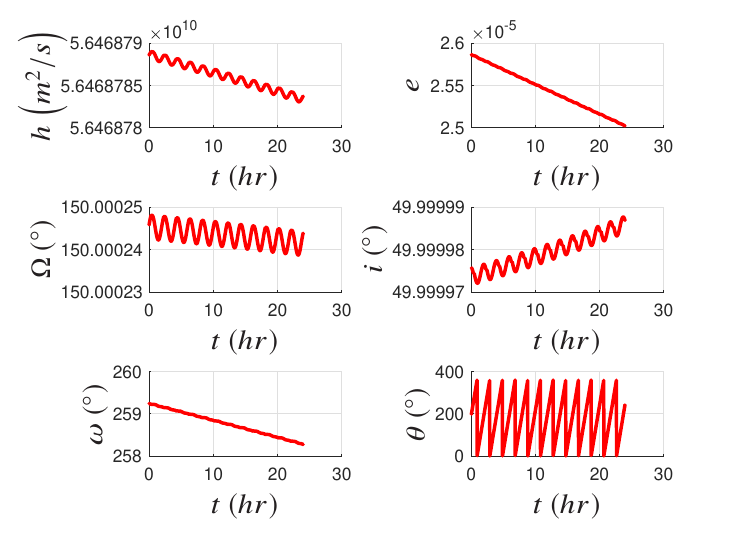}
\label{esttrajleo}}
    \caption{SO Trajectory using True and Estimated LQR Weights for a LEO Orbit}\label{trajplotleo}
\end{centering}
\end{figure}

\begin{figure}[!htb]
\begin{centering}
    \subfigure[True Cost Function]{
    \includegraphics[keepaspectratio, width=0.45\textwidth]{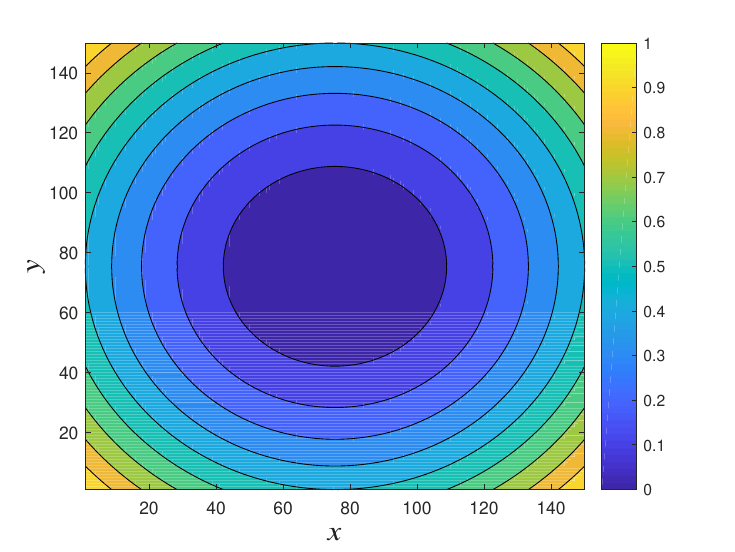}
\label{expertconleo}}
\subfigure[Estimated Cost Function]{
\includegraphics[keepaspectratio, width=0.45\textwidth]{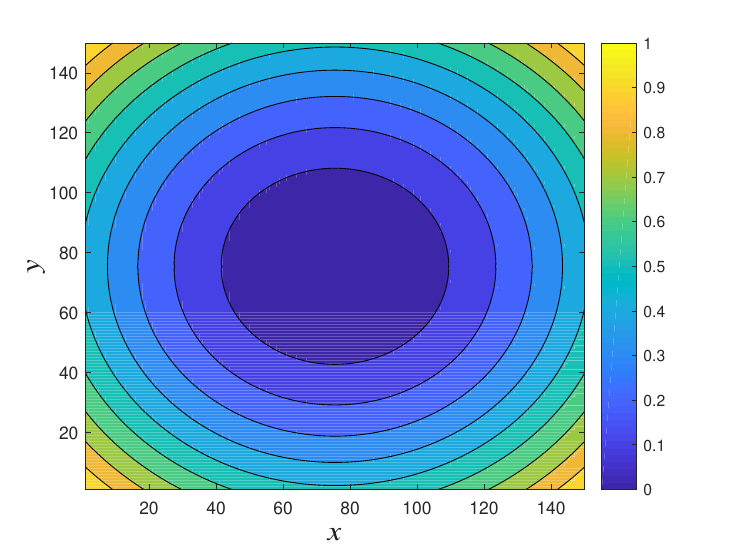}
\label{estconleo}}  
\subfigure[Estimated Cost Function Error]{
\includegraphics[keepaspectratio, width=0.45\textwidth]{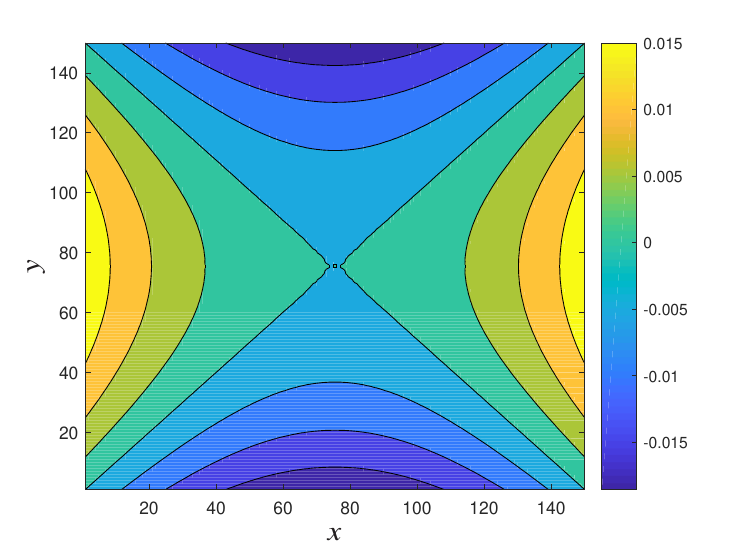}
\label{errorconleo}}  
        \caption{Estimated vs. True Cost Function for LEO Maneuvering SO} \label{fig:case35desLEO}
\end{centering}
\end{figure}

A LEO simulation scenario is also considered for learning the cost function used for SO station-keeping using the parameters in Table \ref{SOparam}. The SO ephemeris for the initial LEO orbit is provided in Table \ref{SOeph}. For the initial circular LEO orbit, the objective is for the SO to stay in LEO under the external disturbances. Similar to the GEO case, the thrusters are fired every two hours for up to 30 minutes to maintain a nominal LEO orbit. One hundred expert trajectories were simulated with Gaussian white noise under constant LQR weight matrices. Figure \ref{experttrajleo} shows a single expert trajectory from an initial perturbation from a LEO orbit. From the figure, the eccentricity, right ascension of the ascending node, and inclination stay approximately constant at $0$, $150^{\circ}$, and $50^{\circ}$, respectively. The oscillations are due to external disturbances on the SO which are not significant enough to perturb the SO outside a $75km$ box from the nominal LEO orbit in a 24 hour period. Thus, no thruster burns for SO station-keeping are observed. Figure \ref{expertconleo} shows the true (and unknown) cost used for every expert trajectory. Since no control thrust is used, this contour does not depend on $R$.

\begin{table}[htbp]
   \caption{SO ephemeris at an initial LEO orbit} \label{SOeph}
   \small 
   \centering 
   \begin{tabular}{lr} 
   \toprule[\heavyrulewidth]\toprule[\heavyrulewidth]
   \textbf{Parameter} & \textbf{Value} \\ 
   \midrule
   Semimajor Axis  & 8000 km \\
   Eccentricity & 0\\
   Inclination & 50$^{\circ}$ \\
   Right Ascension of the Ascending Node & 150$^{\circ}$\\
   Argument of Perigee & 95$^{\circ}$\\
   True Anomaly & 0$^{\circ}$\\
   Epoch & July, 1st, 2020 $00:00:00.0$ UT1\\
   \bottomrule[\heavyrulewidth] 
   \end{tabular}
\end{table}

Figure \ref{esttrajleo} shows a trajectory using the estimated LQR weights found using maximum causal entropy. The estimated trajectory is initialized with the same initial conditions as Figure \ref{experttrajleo}. The trajectory that is determined using the estimated weights has an almost identical response to the expert trajectory. Also, the SO trajectory obtained from the estimated weights did not use any thruster burns to maintain its nominal orbit. Thus, it follows the behavior of the expert. The estimated cost function is shown in Figure \ref{estconleo}. The estimated almost matches the behavior of the expert trajectory. In Figure \ref{errorconleo}, the maximum error is $0.015$ between the true and estimated cost functions. Since the trajectories obtain from the expert and estimated weights uses no control within the 24 hour period, the behavior from the estimated weights captures the general trend of the LEO station-keeping maneuver.

\section{Conclusion}
The objective of the paper is to formulate inverse RL to learn the behavior of SOs from observed orbital motion using maximum causal entropy and provide examples for a SO in GEO and LEO. By setting up expert SO trajectories using an LQR-based cost, the SO behavior is learned using maximum causal entropy inverse RL. It is shown that the learned cost function in LEO or GEO is comparable to the observed expert SO's trajectory in the presence of orbital disturbances. Thus, maximum causual entropy is attractive in learning the reward function that produces the expert SO's maneuver and can be used to estimate the behavior of SOs. 
\section*{Acknowledgement}
This research was supported by an appointment to the Intelligence Community Postdoctoral Research Fellowship Program at Massachusetts Institute of Technology, administered by Oak Ridge Institute for Science and Education through an interagency agreement between the U.S. Department of Energy and the Office of the Director of National Intelligence.
\bibliography{biblio}
\end{document}